\documentclass[twocolumn,aps,prl,groupedaddress,amssymb,showpacs]{revtex4}
\usepackage{graphicx}
\usepackage{dcolumn}
\usepackage{amsmath}

\begin{document}

\title{"Gradient marker" -- a universal wave pattern in inhomogeneous continuum}

\author{A. E. Kaplan}
\email{alexander.kaplan@jhu.edu}
\affiliation{Electr. and Comp. Eng. Dept, The Johns Hopkins University, Baltimore, MD 21218}

\date{\today}

\begin{abstract}
Wave transport in a media with slow spatial gradient 
of its characteristics is found to exhibit a universal
wave pattern ("gradient marker") 
in a vicinity of the maxima/minima of the gradient.
The pattern is common for optics, quantum
mechanics and any other propagation
governed by the same wave equation.
Derived analytically,
it has an elegantly simple
yet nontrivial profile found in perfect 
agreement with numerical simulations for specific examples. 
We also found resonant states
in continuum in the case of quantum
wells, and formulated criterium for their existence.
\end{abstract}


\maketitle

Wave  patterns in inhomogeneous media
or confining structures are of great interest 
to quantum mechanics, optics and electrodynamics,
acoustics, hydrodynamics, and chemistry.
Examples include wave packets in atoms [1],
Ghladny patterns in acoustics, 
EM resonator and waveguide modes [2],
Anderson localization in disordered systems [3],
soliton formation [4] due to nonlinearity,
including atomic solitons in bosonic gas [5], 
as well as giant waves near caustics [6],
waves in chemical reactions [7],
dark-soliton grids [8], "scars" in "quantum billiard" [9],
"quantum carpets" in QM potentials  [10],
nano-stratification of local field in 
finite lattices [11], etc.
In all of those,
the presence of multi-modes or a broad-bend spectrum
is pre-requisite for interference and pattern formation
in inhomogeneous or confining structures.

In this Letter we show, however, that a 
localized wave pattern -- an immobile single-cycle intensity profile --
can emerge in a single-mode wave in a vicinity
of a min/max of the {\emph{gradient}} of QM potential or 
optical refractive index.
The phenomenon is universal for both optics
and quantum mechanics, and for any 
other propagation described by a wave
equation (1) below.
What makes it unusual is that
it emerges in media with \emph {no potential wells}
and only a smooth inhomogeneity yielding no reflection,
-- and is originated by a purely traveling wave
with apparently no other modes to interfere with.
We found, however, that this wave 
here generates a co-traveling but localized "satellite" of
slightly different phase and amplitude
resulting in "self-interference".
The wave ideally is not trapped and carries
its momentum and energy flux unchanged through the area.
To a degree, the pattern  mimics a 2-nd order
spatial derivative of the refractive index (or potential function);
it would be natural to call it a "gradient (G) marker".
In QM it may be most pronounced 
for an above-barrier propagation
of electron in continuum over smoothly-varying
potential; in solid state it might
emerge above the critical temperature
for the Anderson localization to vanish.
Even for a potential well,
when the energy of electrons exceeds 
the ionization potential and there is no trapping, 
the G-markers emerge as the main non-resonant localized feature.

To demonstrate the effect and elucidate analytical results 
(to be compared with numerical simulations)
we consider 1D-case  
written, for the sake of compactness,
in "optical" terms, using space-varying refractive index $n ( x )$;
yet we consistently
"translate" all the effects and approaches into QM-terms.
A 1D spatial dynamics of an $\omega$-monochromatic plane wave 
with linearly polarized electrical 
field $\vec{E} = {\hat{e}}_p  E ( x ) \exp ( - i \omega t ) + c.c.$, 
propagating in the $x$-axis
(here ${\hat{e}}_p \perp {\hat{e}}_x$ is a polarization unity vector),
is governed by wave equation
\begin{equation}
E^{\prime \prime}  +  n^2 ( \xi )  E  = 0 ;
\ \ \ \ \ \xi = x k_0 ; 
\label{(1)}
\end{equation}
where $k_0  = \omega / c = 2 \pi / \lambda_0$,
and "prime" stands for $d / d \xi$.
(For $\vec{H}$ field, $\hat{e}_x \perp \vec{H}  \perp \hat{e}_p$,
one has $H  = - i  E^\prime$ in non-magnetic materials;
for a traveling wave, $|H|  = n |E|$, if $n = const$.)
In QM-terms, this corresponds to 1D-scattering of a particle 
in continuum by a potential $U ( x )$, with 
$E$ replaced by a wave function, $\psi$, of a particle,
$H$ - by $ ( - i \psi^\prime ) = p_Q / k_0 \hbar$, 
$n$ - by  $p_C / k_0 \hbar$, 
where $p_Q$ and $p_C = \sqrt {2 m [ {\bf E_0} - U ( x ) ]}$
are its quantum and classical momenta respectively, 
and $\bf E_0$ -- full energy.
We will consider only the case
$n^2 > 0$, where one can attain a no-reflection
mode of the main interest to us here; 
otherwise, with $n^2 ( \xi )$ crossing zero,
the system may exhibit a full reflection 
characterized by an Airy function
as e. g. near a turning point in QM [12],
or a critical point in plasma [2], 
or caustics in optics and water waves [6].

Eq (1) is ubiquitous in physics and engineering.
Since few known functions $n ( \xi )$  allow for analytical solutions,
numerical simulations and/or 
approximate analytical solutions in general have to be used.
Of the most interest to us here
will be the limit of adiabatically \emph {slow} variation
in space, when gradient parameter 
$\mu \sim ( k_0 L n_{min} )^{-1}$,
where $L$ is a spatial scale 
of inhomogeneity, is small, $ \mu \ll 1$,
which corresponds to a quasi-classical case in QM.
The reflectivity $R$ in this case
vanishes as $R = O ( e^{ - A / \mu }  )$ [12,13],
where $A = O ( 1 )$  (usually $A > 1$),
and reflection can be neglected by a large margin.
A solution is provided then
by a WKB approximation [12] as traveling waves,
$ C^{( \pm )} $ $\exp ( \pm  i \int n d \xi ) / n^{1/2}$.
Considering e. g. a forward wave,
and setting $ n^\prime \rightarrow 0$
at $| \xi | \rightarrow \infty$, where  
we normalize its intensity by setting $n | E {|^2}_{\infty} = 1$,
we look for next approximation as a perturbed WKB solution
\begin{equation}
E = [ 1 + \Delta ( \xi ) ] e^{i \int n d \xi } / n^{1/2}
\ \ \ with \ \ \ \Delta = \gamma + i \beta
\label{(2)}
\end{equation}
where $\Delta ( \xi )$ ($| \Delta |^2 \ll 1 $)
is a slow-varying complex function, 
$\Delta \rightarrow 0$ at $| \xi | \rightarrow \infty$,
$ \gamma $ and $ \beta $ - real;
as we will see later on, $| \gamma |_{max} \sim | \beta^\prime |_{max} 
= O ( \mu^2 )$.  
Using ansatz (2) in Eq. (1),
setting real and imaginary parts of the sum of
all the perturbations terms to zero,
and collecting the terms of lower order in $\mu$
in each one of them, we obtain
for the real part an equation consisting
of $O ( \mu^2 )$ terms
\begin{equation}
\beta^{\prime} = -
( {n^{\prime} / n^{3/2}} )^{\prime} / {4 n^{1/2}}
\label{(3)}
\end{equation}
and for the imaginary part - an equation consisting
of $O ( \mu^3 )$ terms, integration of which yields
$\gamma = - ( \beta^2 + \beta^{\prime} / n ) / 2$,
where we set the integration constant to zero due
to above condition $\Delta_{ \pm \infty} = 0$.
(It is worth noting that in the end,
all the terms with $\beta^2$ get canceled,
so there is no need for further integration of Eq. (3).)
We can finally arrive at a G-marker intensity by
calculating the perturbation, $\delta I ( \xi ) \equiv I - 1$, 
of the normalized
field intensity $ I \equiv n | E |^2 = ( 1 + \gamma )^2 + \beta^2
\approx 1 + 2 \gamma + \beta^2$,
retaining the terms lowest in $\mu$, 
and obtaining to $o ( \mu^2 )$:
\begin{equation}
\delta I ( \xi ) = - {\beta^{\prime} / n} =
( {n^{\prime} / n^{3/2}} )^{\prime} / {4 n^{3/2}}
\label{(4)}
\end{equation}
In the vicinity of a gradient peak,
$\delta I ( \xi )$ makes an asymmetric
single-cycle shape, with its
middle point shifted by $O ( k_0 L )$ toward the area with lower
refractive index (or higher potential);
its higher (and positive at that)
peak is also located in the same area, see Figs. 1 and 2.
One can see that $\delta I ( \xi )$ more or less
mimics a second derivative of $n$.
\begin{figure} 
\begin{center}
\includegraphics[width=5.2cm, height=8.6cm, trim=0 15mm 0 0, clip, angle=270]{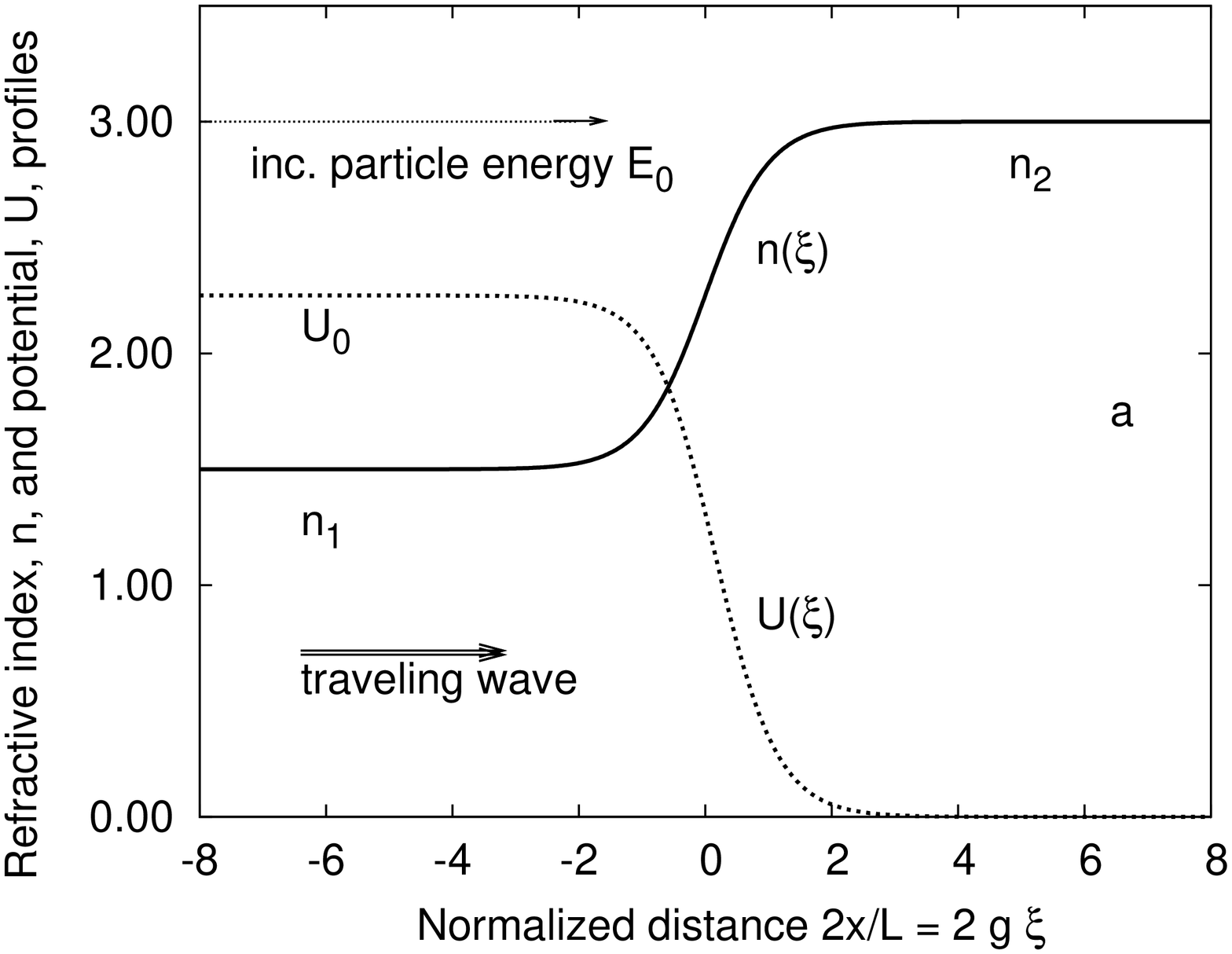} 
\includegraphics[width=5.4cm, height=8.6cm, trim=0 15mm 0 0, clip, angle=270]{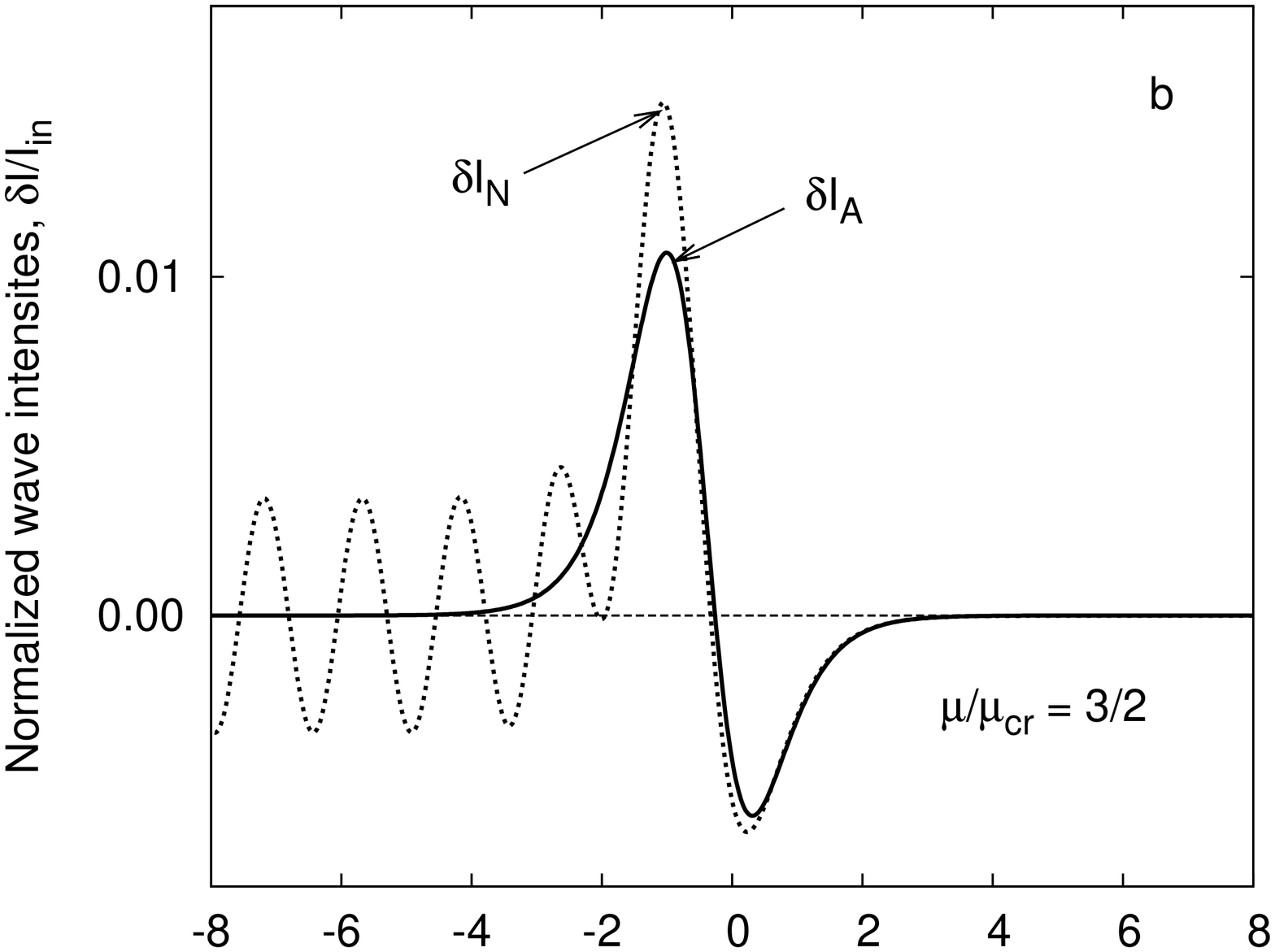}
\includegraphics[width=5.4cm, height=8.6cm, trim=0 15mm 0 0, clip, angle=270]{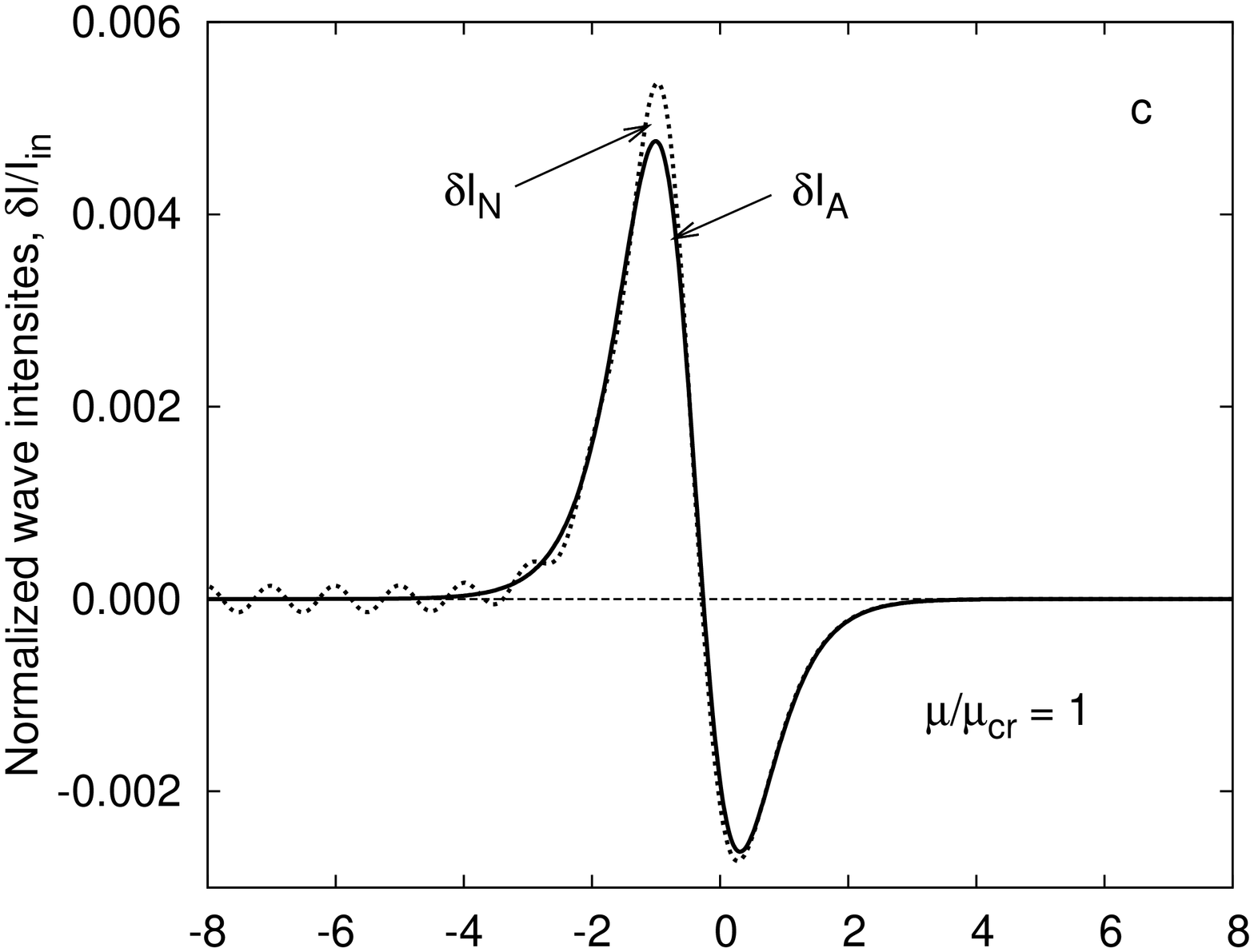}
\includegraphics[width=5.4cm, height=8.6cm, trim=0 15mm 0 0, clip, angle=270]{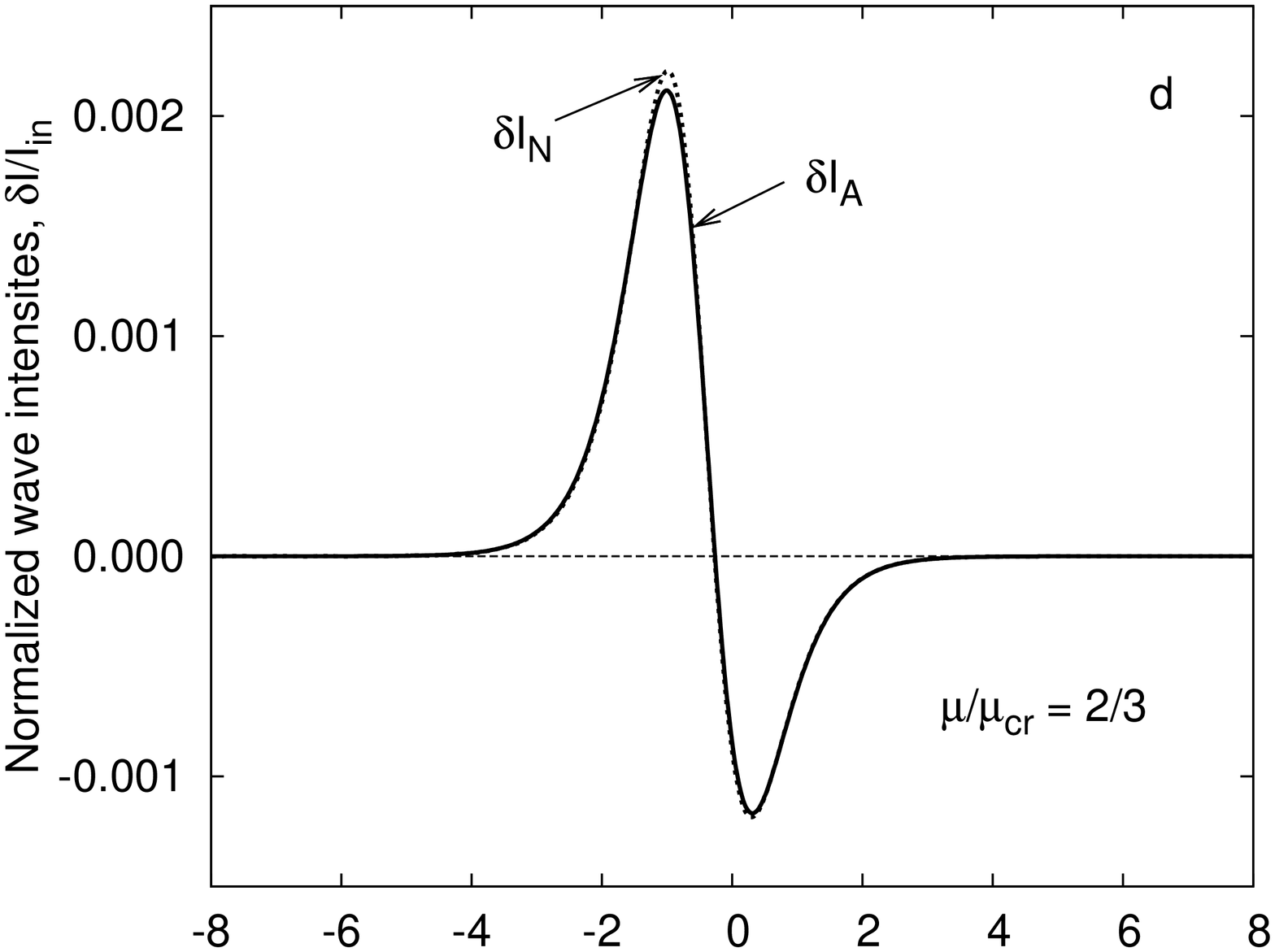}
\caption{(a) Refractive index, $n$ (and potential $U$)
soft-step spatial profiles, $n_1 = 1.5$ and $n_2 = 3$,
$\mu_{cr} \approx 0.24$;
(b-d) G-marker intensity, $\delta I$, {\it {vs}}
distance $x/2L$ for various parameters $\mu$;
curves: $\delta I_N$ - numerical, and $\delta I_A$ -
analytical, Eq. (4).}
\label{fig1}
\end{center}
\end{figure}
Eq.(4) can also be obtained {\it{via}}
quasi-classical approximation in QM [12],
whereby one has to search for high-order
corrections for the phase of $\psi$
as function of the classical momentum $p_C$,
after which it has to be 
translated into correction to intensity.

How far the asymptotic result (4) can be pushed beyond the limit
$\mu \ll 1$, and what
is a critical $\mu_{cr} = O ( 1 )$,
can be explored only by numerical simulations,
which also helps to reveal 
a real nature of a small parameter
$\mu$ (which appears to be substantially different from a standard
$( | n^\prime | / n^2 )_{max} \ll 1$ [14]).
Before comparing Eq. (4) to numerical simulations for specific profiles
$n ( \xi )$ and various $\mu$, let us make sure it conforms
to the conservation of EM energy flux,
i. e. (time-averaged) magnitude, $\overline{S}$, of the Poynting vector, 
$\vec{S} = \vec{E} \times \vec{H} / 2$ (in Abraham's form) 
in general case.
Writing $S = ( E e^{- i \omega t}$ $ + c. c. ) \cdot $
$( H e^{- i \omega t} + c. c. ) / 2 $,
and $t$-averaging it, which amounts here to omitting
terms with $e^{\pm 2 i \omega t}$,
we have $\overline{S} = Re ( E H^* )$ 
$= Re ( i E^{* \prime} E )$.
In QM terms, it corresponds to mathematical expectation
of a particle momentum, $< \psi | \hat{p}_Q | \psi >$,
$\hat{p}_Q = -i \hbar d/dx$.
Using Eq. (2) and retaining the terms 
of the lowest (2-nd here) order in $\mu$,
we have $\overline{S} = I + \beta^{\prime} / n$;
due to Eq. (3) it confirms that $\overline{S} = 1 = inv$
to $o ( \mu^2 )$.

For numerical simulations of Eq. (1) with an arbitrary profile
$n ( \xi )$ and \emph {arbitrary} $\mu$,
we broke it into two 1st order
(Maxwell, in $EM$-case) equations:
$E^{\prime} = i H$; $H^{\prime }= i  n^2  E $.
To model a "soft step" $n ( \xi )$, we use a function (Fig. 1a):
\begin{equation}
n ( x )  =  n_1  +  
( n_2  -  n_1 )  {\bf [} 1 + \tanh ( 2 x / L ) {\bf ]} / 2  
\label{(5)}
\end{equation}
with controllable $L$, $n_i$, and its gradient parameter as
\begin{equation}
\mu = g ({n_1}^{-1} + {n_2}^{-1}) ; \ \ \ \ g = ( k_0 L )^{-1} =
\lambda_0 / 2 \pi L
\label{(6)}
\end{equation}
[14]; the spectral dispersion of $n_i$ can be safely ignored here.
The calculations with an arbitrary $\mu$ are done by using a multi-point
algorithm and a "reverse propagation" mode,
whereby we start at $\xi \gg g^{-1}$, 
postulate that only one (transmitted) field remains there,
$E_{{\infty}} \to \exp ( i n_2 \xi ) / \sqrt n_2$,
(the Sommerfeld's condition), and then go backward, 
till reaching a symmetrically located area in front of the gradient,
$\xi \ll - g^{-1}$,
where we record the intensity of an incident wave,
$I_{in} = n_1 | E + H / n_1 |^2 / 4$,
and then normalize all the stored intensities by $I_{in}$ [15].
The precision of calculations is checked 
against the deviation of
$\overline{S}$, at each point
from that recorded at the incidence; 
typically it was better than $ 10^{-6}$.

The retroreflection from the gradient area is strongly suppressed
and G-marker is well emphasized (see e. g.
Figs. 1c and 1d) provided that $\mu < \mu_{cr}$,
where parameter $\mu_{cr}$ was found by us to be almost universal,
$\mu_{cr} \approx  (2 \pi )^{-1}$ at
$r_n \equiv n_1 / n_2 + n_2 / n_1 \gg 1$,
and slightly increasing to $\mu_{cr} \approx 1/4$ 
near $ n_1 \sim n_2$.
The highest (positive) G-marker peaks, $\delta I_{max}$,
can easily reach a few percent of the intensity, $I$,
especially at $r_n \gg 1$.
In the case of a "shallow"
soft step, $| n_1 - n_2 | \ll \tilde n / 2$, 
$\tilde n = ( n_1 + n_2 ) / 2$
[in QM this would correspond 
to a kinetic energy $\bf E_0$
much higher than the drop of potential, $U_0$,
${\bf {E_0}} / U_0 \sim \tilde n / | n_1 - n_2 | \gg 1$],
the max/min of $\delta I$ are almost of the same magnitude,
\begin{equation}
| \delta I_{M} | \approx 2 g^2 {| n_1 - n_2 | } / ( 5^{3/2} { \tilde n^4} )
\label{(7)}
\end{equation}
and located at $x \approx \pm L/4$.
(In general, the parallel between optics
and QM can be guided by the relationship 
$U_0 / {\bf {E_0}} = 1 - \min [ ( n_1^2 / n_2^2 ) ,  
( n_2^2 / n_1^2 ) ]$.)

Figs. 1b-1d for the case
of $n_1 = 1.5$, $n_2 = 3$, $\mu_{cr} =$ $ 0.24$,
show numerical simulations of spatial dynamics
of $\delta I_N$,  
converging amazingly fast to an asymptotic
analytical result for a G-marker intensity, $\delta I_A$, Eq. (4),
as soon as $\mu \leq \mu_{cr}$.
Fig.1b with $\mu / \mu_{cr} = 3/2$
shows a residual reflection
giving rise to an oscillating structure (partial
standing wave), comparable in its amplitude
to a G-marker,
while Fig. 1b ($\mu = \mu_{cr}$) depicts
distinct and strong G-marker formed even at
$L \approx \lambda_0 /2$.
Finally, an inhomogeneity with $L = \lambda_0$
(Fig. 1d, $\mu / \mu_{cr} = 2/3$) is sufficient
to produce a very clean G-marker.

We move now to investigate G-marker formation
by a potential well (or a refractive index plateau)
by modeling it with an "up-and-down" double step, Fig. 2a:
\begin{equation}
n ( x ) = n_1 + ( n_2 - n_1 ) 
({T_{+} - T_{-}) / { 2 \tanh (D/L) } 
}
\label{(8)}
\end{equation}
where $T_\pm = \tanh [ ( 2 x \pm D )/L ]$,
and $D$ is a controllable spacing between the steps.
For $D \ll L$, it becomes
$n ( x ) = n_1 + ( n_2 - n_1 ) / \cosh^2 (2x/L)$,
but for our 
purposes here we choose more box-like well, $D / L = 8$,
which has $\mu$ defined by Eq. (6), 
and the same $\mu_{cr}$ as a soft-step (5).
\begin{figure} 
\begin{center}
\includegraphics[width=5.2cm, height=8.6cm, trim=0 15mm 0 0, clip, angle=270]{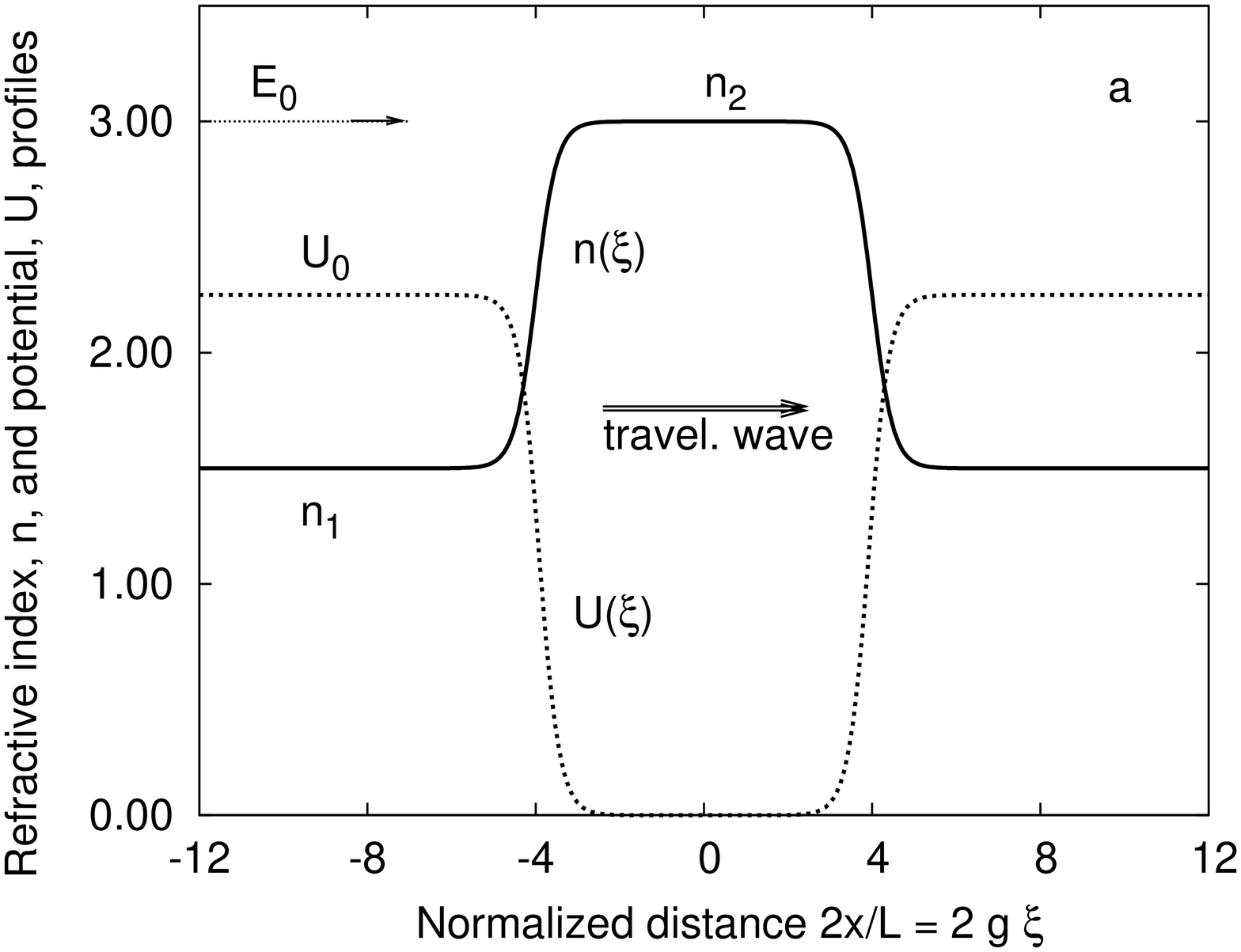}
\includegraphics[width=5.4cm, height=8.6cm, trim=0 15mm 0 0, clip, angle=270]{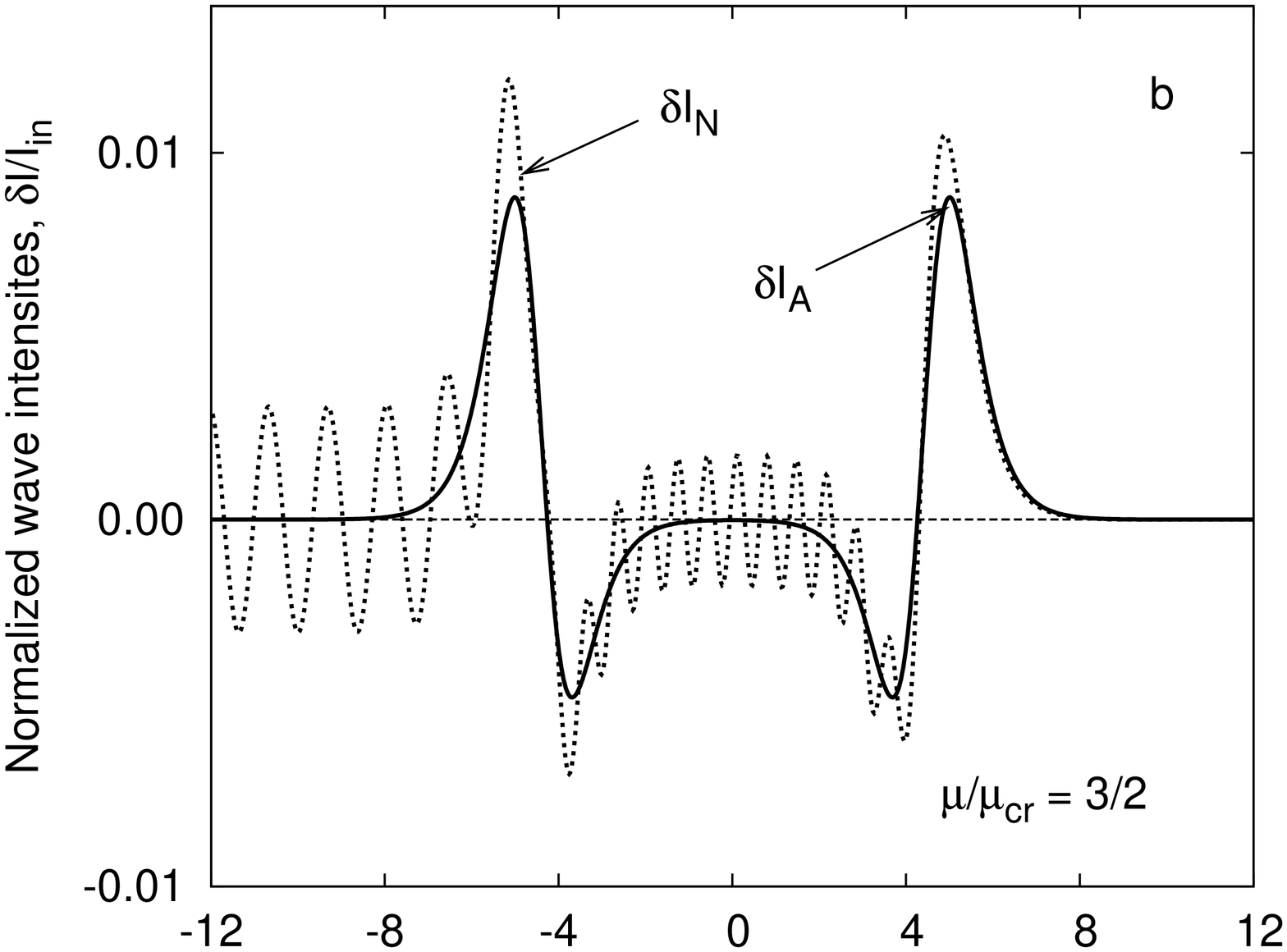}
\includegraphics[width=5.4cm, height=8.6cm, trim=0 15mm 0 0, clip, angle=270]{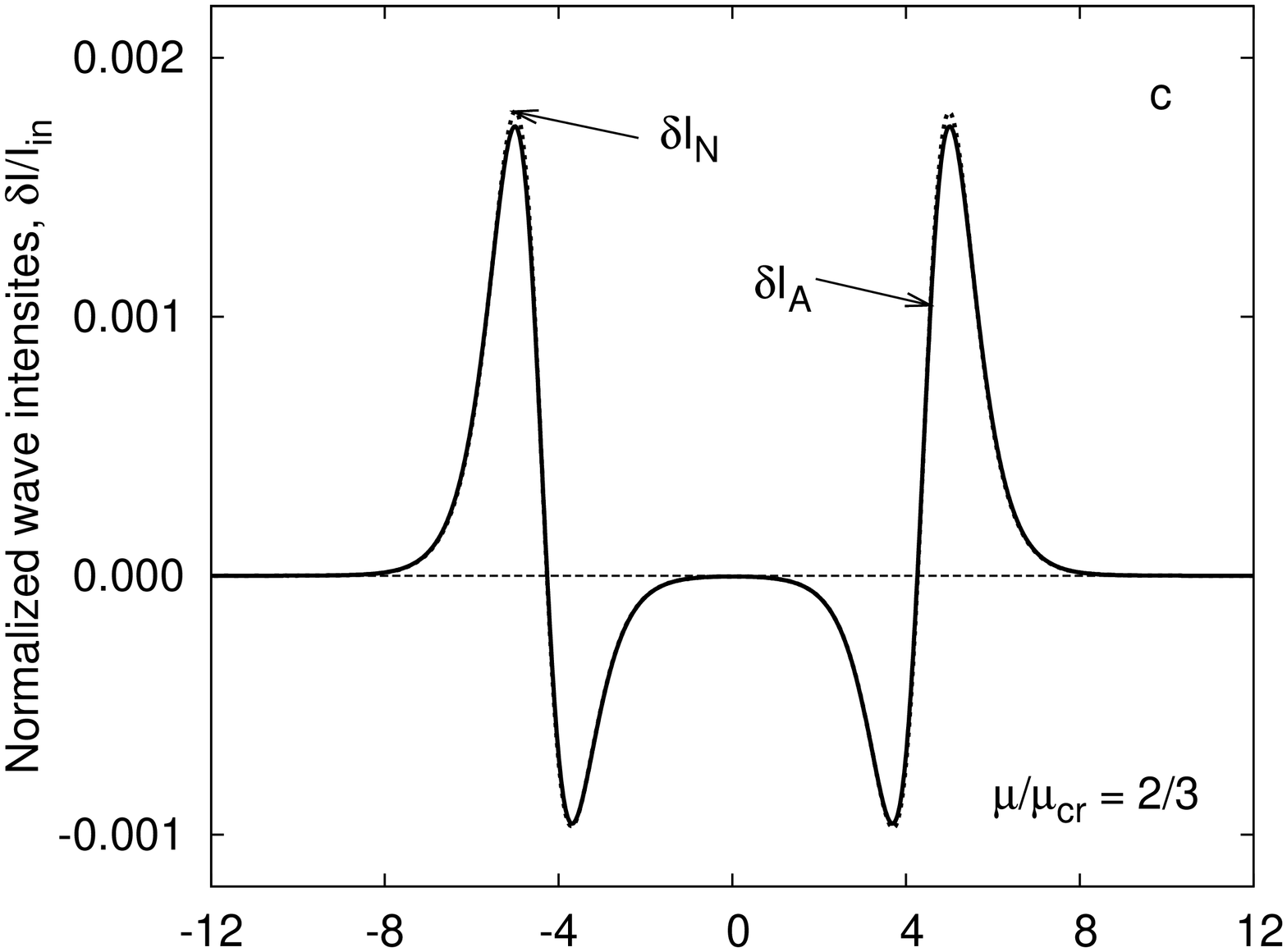}
\includegraphics[width=5.4cm, height=8.6cm, trim=0 15mm 0 0, clip, angle=270]{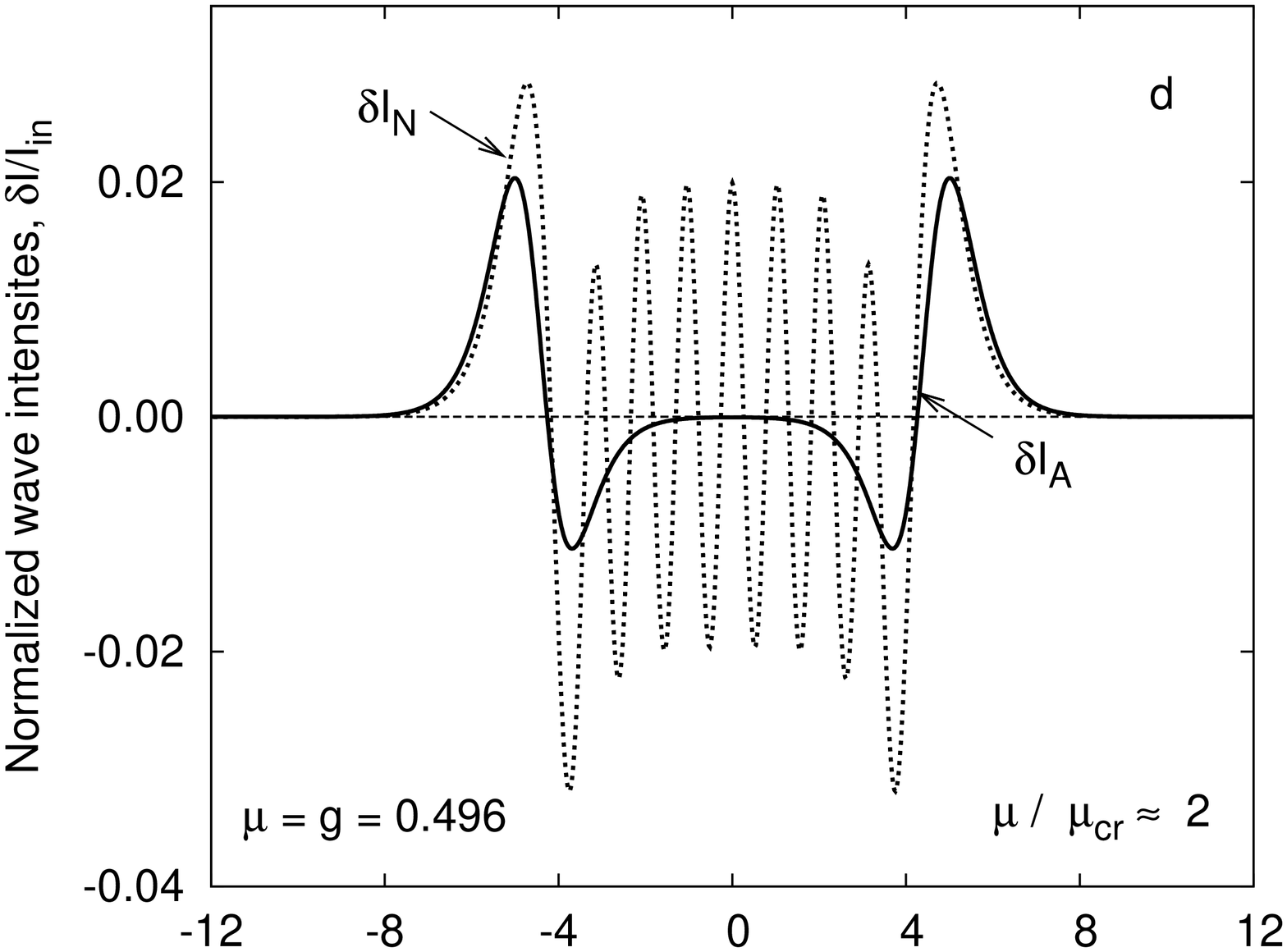}
\caption{(a) Refractive index plateau, $n$ (and potential well $U$)
profiles, with $n_1$, $n_2$, and $\mu_{cr}$ same as in Fig. 1;
(b,c) $\delta I$, {\it {vs}} $x/2L$ for 
$\mu > \mu_{cr}$ (b), and $\mu < \mu_{cr}$ (c),
with $\delta I_N$ and $\delta I_A$ as in Fig.1;
(d) resonant state in continuum (see the text).
}
\label{fig2}
\end{center}
\end{figure}
As expected, both walls form 
G-markers symmetric to each other, Figs. 1b and 1c, so that
to form a G-marker it does not matter
which way a wave is arriving - from the lower
index or from the higher one.
At $\mu > \mu_{cr}$ one can see some
oscillations, same as for a single wall in Fig. 1b
for the same $\mu$, and ideally clean G-markers
for $\mu < \mu_{cr}$, similar to Fig. 1d
for the same $\mu$.

The major difference here
comes, however, in the area $\mu > \mu_{cr}$.
Here, at certain (countable) set of points in the continuum,
while there are strong oscillations
within a potential well, which indicates
a significant wave reflection \emph{between} G-markers,
there is no reflection from
the \emph{entire potential well}, see Fig. 2d.
That confined partially standing wave 
is a signature of a \emph{resonant state}
in a finite-depth quantum well with rigid walls,
most known in the case of a finite rectangular box.
Fig. 2d depicts one of those states with ${\bf{E_0}} / U_0 = 4/3$.
The condition for them
to emerge above a quantum well 
is a significant rigidity of the well's  walls, $\mu > \mu_{cr}$.
In the limit $\mu \gg \mu_{cr}$,
their energies in the continuum 
coincide with those of a  finite box, or in turn - with
the eigenstates of a box with infinitely-high walls,
${\bf E_N} = ( N \hbar \pi )^2 / 2 m D^2$,
where $N$ is a natural number, provided that ${\bf E_N} > U_0$.
In optics terms, they correspond to full-transmission resonances
of a Fabri-Pierrot resonator with semi-transparent mirrors.
In solid-state, these states 
may reveal themselves during a $\delta$-kick field ionization 
\emph{via} production of spatially-stratified bunches in photoelectron
current, whose kinetic energies coincide with those of
the resonant states [16].

Potential uses/applications of 1D (or almost 1D) 
G-markers can be envisioned, such as
(a) observation of quantum "traces" in continuum, i. e.
beyond "quantum carpets" [9] in potential wells,
(b) detection and control of slight changes
of optical fiber parameters [17],
(c) the diagnostics of cold under-dense plasma,
(d) medical surface-wave ultrasound tomography,
(e) detection of the movement of near-shelf
profiles of the bottom of oceans and rivers by 
space- or air-borne photography of the patterns of
wind-driven gravitation waves, as well as 
(f) contour-detection and tracing of submerged 
large/long moving man-made objects or whales in the ocean.

A 2D and 3D expansion of the theory may need to be developed
for other potential applications of G-markers such 
(g) the "tomography" of quantum landscape
in disordered solid-state at above-critical temperature [3],
(i) a \emph{bulk} tomography of opaque fluids (e. g. oil or
muddy water) by using non-penetrating \emph{surface}
EM or acoustic waves, or of solid-state bodies
(e. g. in "introvision" of computer chips, 
or lacunas in blobs of metallic alloys or glass), as well as
(j) in plasma- and astro-physics.

In conclusion, we predicted the formation
of a universal feature in wave transport
in a inhomogeneous media  -- a standing
single-cycle spatial modulation
of wave intensity -- gradient marker --
located in closed vicinity of max/min of a gradient
of refractive index or potential function.
We found a critical condition
for a G-marker to be resolved on the background of residual reflection.
In the presence of a trapping potential,
we also found resonant states/modes
in the continuum at the energies above
photo-ionization and formulated the condition
of those modes to exist when G-markers are not dominant.

This work is supported by AFOSR.

\end{document}